\documentclass[11pt]{article}
\usepackage{amssymb, amsmath, amsthm }
\usepackage{graphicx}
\usepackage{setspace, bbm}
\usepackage{caption}
\usepackage{subcaption}
\newcommand{\ba}{\begin{align*}}
\newcommand{\eaa}{\end{align*}}
\newcommand {\nt} {\notag}
\newcommand{\nl}{\notag\\}

\newcommand{\frn}{\frac 1 n}

\newcommand{\calX}{{\cal X}}

\newcommand{\calY}{{\cal Y}}
\newcommand{\calZ}{{\cal Z}}

\newcommand{\calA}{{\cal A}}
\newcommand{\calB}{{\cal B}}

\newcommand{\calF}{{\cal F}}
\newcommand{\calI}{{\cal I}}
\newcommand{\calS}{{\cal S}}

\newcommand{\calC}{{\cal C}}
\newcommand{\calR}{{\cal R}}

\newcommand {\bX} {\mbox{\boldmath $X$}}

\newcommand {\bZ} {\mbox{\boldmath $Z$}}

\newcommand {\bU} {\mbox{\boldmath $U$}}

\newcommand {\bb} {\mbox{\boldmath $b$}}
\newcommand {\bB} {\mbox{\boldmath $B$}}

\newcommand {\bE} {\mbox{\boldmath $E$}}

\newcommand {\hX}{\hat{X}}

\newcommand {\hx}{\hat{x}}

\newcommand {\hZ}{\hat{Z}}

\newcommand{\eqd}{\stackrel{\triangle}{=}}

\newcommand{\ind}[1] {\mathbbm{1}\left\{#1\right\}}
\newcommand{\sbr}[1] {\left[#1\right]}
\newcommand{\cbr}[1] {\left\{#1\right\}}
\newcommand{\rbr}[1] {\left(#1\right)}
\newtheorem{theorem}{Theorem}

\newtheorem{mydef}{Definition}
\newtheorem{proposition}{Proposition}

\begin{document}
\title{ Zero-Delay and Causal Secure Source Coding\footnote{This research was supported by the Israeli Science Foundation (ISF), grant no. 208/08.}}

\author{Yonatan Kaspi and Neri Merhav\\
Department of Electrical Engineering \\
Technion - Israel Institute of Technology \\
Technion City, Haifa 32000, Israel\\
Email: \{kaspi@tx, merhav@ee\}.technion.ac.il}

%\address{Department of Electrical Engineering \\
%Technion - Israel Institute of Technology \\
%Technion City, Haifa 32000, Israel\\
%Email: \{kaspi@tx, merhav@ee\}.technion.ac.il}

\maketitle
\vspace{-20pt}
\begin{abstract}
We investigate the combination between causal/zero-delay source coding and information-theoretic secrecy. Two source coding models with secrecy constraints are considered. 
We start by considering zero-delay perfectly secret lossless transmission of a memoryless source. We derive bounds on the key rate and coding rate needed for perfect zero-delay secrecy. In this setting, we consider two models which differ by the ability of the eavesdropper to parse the bit-stream passing from the encoder to the legitimate decoder into separate messages.
We also consider causal source coding with a fidelity criterion and side information at the decoder and the eavesdropper. Unlike the zero-delay setting where variable-length coding is traditionally used but might leak information on the source through the length of the codewords, in this setting, since delay is allowed, block coding is possible. We show that in this setting, separation of encryption and causal source coding is optimal.

%We consider zero-delay single-user and multi-user lossy source coding with decoder side information. The zero-delay constraint translates into causal (sequential) encoder and decoder pairs as well as the use of instantaneous codes. For the single-user setting, we show that optimal performance is attained by time sharing at most two scalar encoder-decoder pairs, that use zero-error side information codes. Furthermore, we show that if either the encoder or decoder are restricted a-priori to be scalar, the performance loss compared to an unrestricted system cannot be compensated by the other component and the other component can be scalar as well, even if the zero-delay constraint is lifted. We also demonstrate that, at least in some cases, there is no performance loss (compared to classical arbitrary delay system) from the restriction to sequential decoders.  
%Finally, we show that the multi-terminal source coding problem can be solved in the zero-delay regime and the rate-distortion region is given.
\end{abstract}

\section{Introduction}
We consider several source coding problems with secrecy constraints, in which an encoder, referred to as Alice, transmits the output of a memoryless source to a decoder, referred to as Bob. The communication between Alice and Bob is intercepted by an eavesdropper, referred to as Eve. A secret key is shared between Alice and Bob with which they can respectively encrypt and decrypt the transmission. Attention is restricted to zero-delay and causal settings. Our setting represents time-critical applications, like live multimedia streaming, which need to be transmitted or stored securely so that the contents can be revealed only to authorized parties. Although there is vast literature dealing with source coding with secrecy constraints, as well as works that deal with source coding with delay or causality constraints, very little attention was given in the information theory literature to the combination of those problem areas. 

This paper has two main sections. We start with zero-delay source coding and include secrecy constraints. Our goal is to characterize the pairs of coding rate and key rate (to be formally defined later) with which perfectly secure, zero-delay, lossless transmission is possible. Two models of eavesdroppers are considered, which differ in their ability to parse the bit-stream which is transmitted from Alice to Bob. We continue with the causal source coding setting, as defined by Neuhoff and Gilbert \cite{NeuhoffGilbert1982}, in which delay is allowed, but the cascade of encoder and decoder must be a causal function of the source. In this setting, our goal is to characterize the achievable region of the quadruple composed of rate, distortion, key rate and uncertainty at the eavesdropper (equivocation, formally defined later). This setting is later extended to the scenario where side information (SI),  correlated to the source, is available to Bob and Eve. We introduce each of these settings in more depth and discuss the fundamental difference between them when secrecy is involved in the sequel after reviewing relevant past work. 

Shannon \cite{Shannon1949} introduced the information-theoretic notion of secrecy, where secrecy is measured through the remaining uncertainty about the message at the eavesdropper. This information-theoretic approach of secrecy allows to consider secrecy issues at the physical layer, and ensures unconditionally (regardless of the eavesdroppers computing power and time) secure schemes, since it only relies on the statistical properties of the system. Wyner introduced the wiretap channel in \cite{Wyner1975} and showed that it is possible to send information at a positive rate with perfect secrecy as long as Eve's channel is a degraded version of the channel to Bob. When the channels are clean, two approaches can be found in the literature of secure communication. The first assumes that both Alice and Bob agree on a secret key prior to the transmission of the source. The second approach assumes that Bob and Eve (and possibly Alice) have different versions of SI and secrecy is achieved through this difference.

For the case of shared secret key, Shannon showed that in order for the transmission of a DMS to be fully secure, the rate of the key must be at least as large as the entropy of the source. Yamamoto (\cite{Yamamoto1997} and references therein) studied various secure source coding scenarios that include an extension of Shannon's result to combine secrecy with rate--distortion theory. In both \cite{Shannon1949},\cite{Yamamoto1997}, when no SI is available, it was shown that separation is optimal. Namely, using a source code followed by encryption with the shared key is optimal. The other approach was treated more recently by Prabhakaran and Ramchandran \cite{Prab-Ramch1997} who considered lossless source coding with SI at both Bob and Eve when there is no rate constraint between Alice and Bob. It was shown that the Slepian-Wolf \cite{SlepianWolf73} scheme is not necessarily optimal when the SI structure is not degraded. Coded SI at Bob and SI at Alice was considered in \cite{GunduzEkripPoor2008}. These works were extended by Villard and Piantanida \cite{VillardPaint2011} to the case where distortion is allowed and coded SI is available to Bob. Merhav combined the two approaches with the wire--tap channel \cite{Merhav2008}. In \cite{SchielerCuff2013}, Schieler and Cuff considered the tradeoff between rate, key rate and distortions at the eavesdropper and  legitimate decoder when the eavesdropper had causal access to the source and/or correlated data.  Note that we mentioned only a small sample of the vast literature on this subject. In the works mentioned above, there were no constraints on the delay and/or causality of the system. As a result, the coding theorems of the above works introduced arbitrary long delay and exponential complexity.

The practical need for fast and efficient encryption algorithms for military and commercial applications along with theoretical advances of the cryptology community, led to the development of efficient encryption algorithms and standards which rely on relatively short keys. However, the security of these algorithms depend on computational complexity and the intractability assumption of some hard problems. 
From the information-theoretic perspective, very little work has been done on the intersection of zero-delay or causal source coding and secrecy. A notable exception is \cite{UduwerelleHoISIT2012} which considered the combination of prefix coding and secrecy\footnote{\cite{UduwerelleHoISIT2012} was presented in parallel to the ISIT2012 presentation of this paper \cite{Me-ISIT2012}.}. The figure of merit in \cite{UduwerelleHoISIT2012} is the \textit{expected key consumption} where it is assumed that Alice and Bob can agree on new key bits and use them during the transmission. However, no concrete scheme was given on how these bits will be generated securely. 

For causal source coding, it was shown in \cite{NeuhoffGilbert1982}, that for a discrete memoryless source (DMS), the optimal causal encoder consists of time--sharing between no more than two memoryless quantizers, followed by entropy coding. In \cite{TsachyNeri2005}, Weissman and Merhav extended \cite{NeuhoffGilbert1982} to include SI at the decoder, encoder or both. The discussion in \cite{TsachyNeri2005} was restricted, however, only to settings where the encoder and decoder could agree on the reconstruction symbol. In \cite{Me-RTRD2013} this restriction was dropped. Zero-delay source coding with SI for both single user and multi-user was also considered in \cite{Me-RTRD2013}. 

Without secrecy constraints, the extension of \cite{NeuhoffGilbert1982} to the zero-delay case is straightforward and is done by replacing the block entropy coding by instantaneous Huffman coding. The resulting bit-stream between the encoder and decoder is composed of the Huffman codewords. However, this cannot be done when secrecy is involved, even if only lossless compression is considered. To see why, consider the case where Eve intercepts a Huffman codeword and further assume the bits of the codeword are encrypted with a one-time pad. While the intercepted bits give no information on the encoded symbol (since they are independent of it after the encryption), the number of intercepted bits leaks information on the source symbol. For example, if the codeword is short, Eve knows that the encrypted symbol is one with a high probability. This suggests that in order to achieve perfect secrecy, the lengths of the codewords emitted by the encoder should be independent of the source and that a simple separation scheme as described above for the causal setting will not achieve perfect secrecy. 

In the last example, we assumed that Eve is informed on how to parse the bit-stream into separate codewords. This will be the case, for example, when each codeword is transmitted as a packet over a network and the packets are intercepted by Eve. Even if the bits are meaningless to Eve, she still knows the number of bits in each packet. Our first contribution in this paper is to show that, albeit the above example, for the class of encoders we consider, the key rate can be as low as the average Huffman codeword length of the source. In contrast to the works mentioned above, our results here are not asymptotic. Note that the length of the transmission was not an issue in the previous works mentioned since block coding (with fixed and known length) was used. 

Our second contribution is the analysis of the case where Eve does not have parsing information and cannot parse the bit-stream into the separate codewords. This will be the case, for example, if Eve acquires only the whole bit-stream, not necessarily in real--time, without the log of the network traffic. Alternatively, it acquires an encrypted file after it was saved to the disk. In this case, when we assume that the length of transmission becomes large, we show that the best achievable rates of both the key and the transmission are given by the Huffman length of the source. In contrast to the results described in the previous paragraph, the results in this scenario are asymptotic in the sense that the probability that the system is not secure vanishes when the transmission length is infinite. 

In the last part of this work we revisit the causal setting of \cite{NeuhoffGilbert1982} and include secrecy constraints. The entropy coding phase of Neuhoff and Gilbert \cite{NeuhoffGilbert1982} makes their scheme impractical. However, their result lower bounds the optimal zero-delay coding performance and gives us insight on how the functional causality restriction affect the rate as compared to classical rate-distortion results (which employ non-causal quantization). 
Following this, we explore how the causality restriction affects the region derived by Yamamoto \cite{Yamamoto1997}. Clearly, to achieve secrecy, one can separate the encryption and source coding. Namely, encrypt the output of the entropy coder in the cascade of \cite{NeuhoffGilbert1982} with the key and decrypt before the entropy decoder at the receiver side. The last contribution of this paper is a proof that the cascade of \cite{NeuhoffGilbert1982} is still optimal and the described separation scheme is optimal with the secrecy constraints. When SI is available, as shown in \cite{Me-RTRD2013}, the cascade of quatizer and entropy coding is no longer optimal. We show, however, that in this case as well, separation between source coding and encryption is optimal. While the direct part of the proof is trivial, since we use separation, the proof of the converse turns out to be quite involved.   

The remainder of the paper is organized as follows: In Section \ref{Sec:RealTimeSecrecy} we deal with the zero-delay setting.  Section \ref{Sec:CausalRD} deals with the causal setting where delay is allowed. Each section starts with a formal definition of the setting. We conclude this work and point to several possible future directions in Section \ref{Sec:Conclusions}.

\section{Zero-Delay Perfect Secrecy}\label{Sec:RealTimeSecrecy}
In this section, we consider the combination of zero-delay source coding and secrecy. The difference between zero-delay and causal source coding boils down to the replacement of the entropy coding in the scheme of Neuhoff and Gilbert \cite{NeuhoffGilbert1982} with instantaneous coding. However, as the example given in the Introduction indicates, replacing the entropy coding with instantaneous coding will result in information leakage to Eve, at least when Eve can parse the bit-stream and knows the length of each codeword. In the setup we consider in this section, neither Bob nor Eve have access to SI. The extension to the case where both Bob and Eve have the same SI is straightforward and will be discussed in the sequel. 
We start with notation and formal setting in Section \ref{Sec:ZDPrelim}. Section \ref{Sec:ZDParsing} will deal with the setting where Eve has parsing information while Section \ref{Sec:ZD_NoParsing} will deal with the setting where Eve cannot parse the bit-stream.

\subsection{Preliminaries}\label{Sec:ZDPrelim}
We begin with notation conventions. Capital letters represent scalar random variables (RV's), specific
realizations of them are denoted by the corresponding lower case letters and their alphabets -- by calligraphic letters. For $i < j$ ($i$, $j$ - positive integers), $x^j_i$ will denote the vector $(x_i,\ldots, x_j)$, where for $i = 1$ the subscript will be omitted. We denote the expectation of a random variable $X$ by $\bE(X)$ and its entropy by $H(X)$.
For two random variables $X,Y$, with finite alphabets $\calX,\calY$, respectively and joint probability distribution $P(x,y)$, the average instantaneous codeword length of $X$ conditioned on $Y=y$ will be given by
\begin{align}
    L(X|Y=y) \eqd \min_{l(\cdot)\in\calA_{\calX}}\left\{\sum_{x\in\calX}P(x|y)l(x) \right\}. \label{eq:Huffman}
\end{align}
where $\calA_{\calX}$ is the set of all possible length functions $l:\calX\to \mathbb{Z}^+$ that satisfy Kraft's inequality for alphabet of size $|\calX|$.
$L(X|Y=y)$ is obtained by designing a Huffman code for the probability distribution $P(x|y)$.
With the same abuse of notation common for entropy, we let $L(X|Y)$ denote the expectation of $L(X|Y=y)$ with respect to the randomness of $Y$. The average Huffman codeword length of $X$ is given by $L(X)$ and will be referred to as the Huffman length of $X$. For three random variables $X,Y,Z$ jointly distributed according to $P(x,y,z)$ We let $L(X|y,Z)$ stand for
\begin{align}
L(X|y,Z)\eqd L(X|Y=y,Z) = \sum_{z}P(z|y) L(X|Y=y,Z=z),\label{eq:CondHuffman}
\end{align}
namely, the average Huffman length of $X$ conditioned on $(Y=y,Z)$ where the expectation is taken with respect to the randomness of $Z$ conditioned on the event that $Y=y$.
Since more information at both the encoder and decoder (represented by the conditioning on $Y$) cannot increase the optimal coding rate, we have that $L(X|Y)\leq L(X)$, i.e., conditioning reduces the Huffman length of the source (this can also be easily shown directly from \eqref{eq:Huffman}). 

We consider the following zero-delay source coding problem: Alice wishes to losslessly transmit the output of a DMS $X$, distributed according to $P(x)$, to Bob. The communication between Alice and Bob is intercepted by Eve. Alice and Bob operate without delay. When Alice observes $X_t$ she encodes it, possibly using previous source symbols, by an instantaneous code and transmits the codeword to Bob through a clean digital channel. Bob decodes the codeword and reproduces  $X_t$. A communication stage is defined to start when the source emits $X_t$ and ends when Bob reproduces $X_t$, i.e., Bob cannot use future transmissions to calculate $X_t$. We will assume that both Alice and Bob share access to a completely random binary sequence, $\bU=(U_1,U_2,\ldots)$, which is independent of the source and will be used as key bits. In addition, Alice has access, at each stage, to a private source of randomness $\cbr{V_t}$, which is i.i.d and independent of the source and the key. Neither Bob nor Eve have access to $\cbr{V_t}$.
Let $m_1,m_2,\ldots,m_n$, $m_i\in\mathbbm{N}$ be a non-decreasing sequence of positive integers. At stage $t$, Alice uses $l(K_t)\eqd m_{t}-m_{t-1}$ bits that were not used so far from the key sequence. Let $K_t\eqd (U_{m_{t-1}+1},\ldots,U_{m_{t}})$ denote the stage $t$ key. The parsing of the key sequence up to stage $t$ should be the same for Alice and Bob to maintain synchronization. This can be done through a predefined protocol where the key lengths are fixed in advance or ``on the fly'' through the transmitted data by a pre-defined protocol (such as detecting a valid codeword). Note that such a scheme can introduce dependencies between the used keys $\{K_t\}$ and the data at the encoder ($X^t,K^{t-1},V^t$). Therefore, while $\bU$ is independent of all other variables, $\{K^t\}$ is not and might depend on the encoder's data through the number of bits from $\bU$ it contains. We define the key rate to be  $R_k= \limsup_{n\to\infty}\frn\sum_{t=1}^n \bE l(K_t)$. 
Let $\calZ$ be the set of all finite length binary strings. Denote Alice's output at stage $t$ by $Z_t\in\calZ$ and let $B_t$ denote the unparsed sequence, containing $l(B_t)$ bits, that were transmitted until the end of stage $t$ (note that $B_t$ does not contain parsing information and therefore is different from $Z^t$). The rate of the encoder is defined by $R\eqd  \limsup_{n\to\infty}\frn\bE l(B_n)$.
When we write $Z_t=f(K_t,A,B,C)$ for some encoding function $f$ and any random variables $(A,B,C)$, this will mean that the length of the key,  $K_t$, depend only on $A,B,C$ and not other variables which are not parameters of $f$.

\begin{figure}[htp]
\centering
\includegraphics[width=0.9\textwidth]{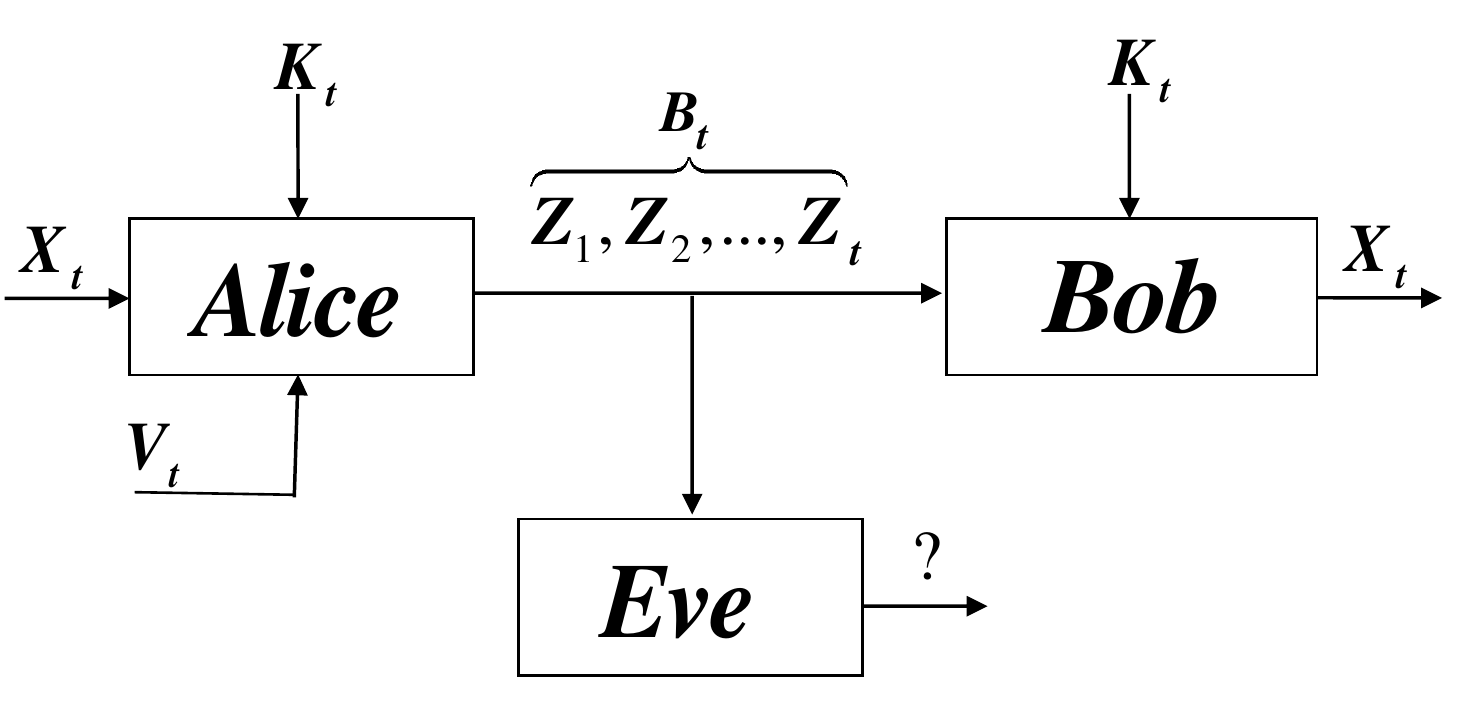} 
\caption{Zero-delay secrecy model}\label{Fig:Secrecy_Model}
\end{figure}

Given the keys up to stage $t$, $K^t$, Bob can parse $B_n$ into $Z_1,\ldots,Z_t$ for any $n\geq t$. The legitimate decoder is thus a sequence functions $X_t=g_t(K^t,Z^t)$. The model of the system is depicted in Fig. \ref{Fig:Secrecy_Model}. 

As discussed in the Introduction, we will treat two secrecy models. In the first model we will assume that Eve can detect when each stage starts, i.e., it can parse $B_t$ into $Z_1,\ldots,Z_t$. In the second model, we will assume that Eve taps into the bit-stream of a continuous transmission between Alice and Bob, $\bB\eqd B_{1}^{\infty}$  but has no information on actual parsing of $\bB$ into the actual messages, $\{Z_t\}$. 
We treat each of the models in detail in the following subsections. 

\subsection{Eve Has Parsing Information} \label{Sec:ZDParsing}
In this subsection, we assume that Eve can parse $B_n$ into $Z_1,Z_2,\ldots,Z_n$. 
In order for the system to be fully secure, we follow Shannon \cite{Shannon1949} in defining what is a secure system. 
\begin{mydef} \label{def_RTSecure}
 	When Eve has parsing information, a system is said to be perfectly secured if for any $l,k,m,n$
	\begin{align}
		P(x^k_l|z_m^n)=P(x^k_l).
 	\end{align}
\end{mydef}
Namely, acquiring any portion of the transmission leaks no information on any portion of the source, which was not known to Eve in advance.

Parsing information is usually not part of any source coding model (with the exception of ``one shot'' models or one-to-one codes \cite{AlonOrlitsky1994}). In this subsection, however, we assume that Eve has this knowledge. It is important to note that this is \textit{additional} information (or side information) that is given to Eve and, unless it is known in advance (for example if fixed length codes are used) it generally cannot be deterministically calculated by observing the encrypted bit-stream which passes from Alice to Bob. The motivation, as discussed in the Introduction, is that sometimes such side information can be obtained by Eve through other means, for example a packet sniffer in a packet network. Moreover, it makes sense to assume that if Eve has such side information, Bob can acquire it as well (for example Eve gets this information through a packet sniffer, obviously, Bob will have this information as well). It is well known that when a decoder has parsing information, there is no need to use uniquely decodable (UD) codes and the average rate of the encoder can be lower than the entropy of the source \cite{AlonOrlitsky1994}. However, it can be easily seen that whenever Eve has parsing information, the rate is lower bounded by $\lceil\log|\calX|\rceil$. 
\begin{proposition}\label{prop:block}
When Eve has parsing information, if the system is perfectly secured then 
\begin{align}
R\geq \lceil\log|\calX|\rceil.
\end{align}
\end{proposition}
\textit{Proof:} Since the number of bits composing $Z_t$ must be independent of the source symbols (otherwise the length of $Z_t$ will leak information, as in the example in the Introduction) the parsing information will assist Bob/Eve to parse the bitstream into $Z_1,Z_2,\ldots$, but will not assist in interpreting the code within $Z_t$. Therefore, within each $Z_t$ an instantaneous code must be used (because of the zero delay constraint). By Kraft's inequality, the shortest block that can accommodate a UD code is of length $\lceil\log|\calX|\rceil$ since
\begin{align}
	1\geq \sum_{x}2^{-l(x)}\geq \sum_{x}2^{-l_{max}}=|\calX|2^{-l_{max}}
\end{align}
where $l(x)$ is the length of the codeword of $x\in\calX$ and $l_{max}$ is the longest codeword in the code. 

The pessimistic conclusion from Proposition \ref{prop:block} is that there is no hope for compression in this setting. As we will see in the following, although such a rate is possible, it will be higher than that, when we will want to minimize the key rate.

The most general zero-delay encoder is a sequence of functions $Z_t = f_t(K^t,V^t,X^t)$. In this section, we will treat only a subclass of encoders that satisfy the Markov chain
\begin{align}
	X_t\leftrightarrow Z^t\leftrightarrow K^{t-1} \label{eq:RT_Markov}.
\end{align}
Namely, given the past and current encoder outputs, the current source symbol, $X_t$, does not reduce the uncertainty regarding the past keys. Similarly, knowing past keys will not assist a cryptanalyst in decrypting the current source symbol. Namely, the past keys are either not reused, and if reused, the resulting message is encrypted by new key bits. 
We claim that this constraint, in the framework of perfect secrecy with a memoryless source is relatively benign and, in fact,  any encoder that calculates a codeword (possibly using the whole history of the source and keys, i.e., with the most general encoder structure), say $\hZ_t$, and then outputs $Z_t=\hZ_t \oplus K_t$ will satisfy this constraint. Such a structure seems natural for one--time pad encryption. In fact, it turns out that \eqref{eq:RT_Markov} includes a broad class of encoders and any \textit{secure} encoder with the general structure of $Z_t=f_t(K_t,V_t,X^t,Z^{t-1})$ satisfies \eqref{eq:RT_Markov}. We show this in Appendix \ref{App:MArkov}. As we will see in the following, although \eqref{eq:RT_Markov} allows for encoder/decoder which can try to use previously sent source symbols which at stage $t$ are known only to Alice and Bob to reduce the number of key bits used at stage $t$, this will not be possible. 

The main result of this subsection is the following theorem:
\begin{theorem} \label{Thm:RT}
 	In the class of codes that satisfy \eqref{eq:RT_Markov}, there exists a pair of perfectly secure zero-delay encoder and decoder if and only if $R_k \geq L(X)$.
\end{theorem}
\noindent\textbf{Remarks:}\\
1) This theorem is in the spirit of the result of \cite{Shannon1949}, where the entropy is replaced by the Huffman length due to the zero-delay constraint. As discussed in the Introduction, variable-rate coding is not an option when we want the communication to be perfectly secret. This means that the encoder should either output constant length (short) blocks or have the transmission length independent of the source symbol in some other way. While by Proposition \ref{prop:block}, no compression is possible, Theorem \ref{Thm:RT} shows that the rate of the key can be as low as $L(X)$ which is the minimal length for zero delay coding. In the proof of the direct part of Theorem \ref{Thm:RT}, we show that a constant rate encoder with block length corresponding to the longest Huffman codeword achieves this key rate. The padding is done by random bits from the encoder's private source of randomness.  

\noindent 2) When Both Bob and Eve has SI, say $Y_t$ where $(X_t,Y_t)$ are drawn from a memoryless source emitting pairs, a theorem in the spirit of Theorem \ref{Thm:RT} can be derived by replacing the Huffman codes by zero-delay zero-error SI aware codes, which were derived by Alon and Orlitsky in \cite{AlonOrlitski96}. These codes also have the property that conditioning reduces their expected length and therefore the proof will contain the same arguments when \eqref{eq:RT_Markov} will be written as $X_t\leftrightarrow (Z^t,Y^t)\leftrightarrow K^{t-1}$ (see also \cite{Me-RTRD2013} for example of these proof techniques using the codes of \cite{AlonOrlitski96}).

\noindent 3) Extending Theorem \ref{Thm:RT} from lossless to lossy source coding is possible by replacing $X$ by $\hX$ where $\hX$ will be the output of a zero-delay reproduction coder. If the distortion constraint is ``per-letter'', then this is straight forward. If the distortion constraint is on the whole block (as in classical rate-distortion) then \eqref{eq:RT_Markov} will impose a strong restriction on the reproduction coder, forcing it to be memoryless (otherwise, knowing past symbols by using $K^{t-1}$ will leak information of the current symbol) . Although zero-delay rate distortion results suggest that indeed $\hX$ should be memoryless (see \cite{NeuhoffGilbert1982}, \cite{Me-RTRD2013}), these results do not restrict the reproduction coder to be memoryless in advance.

We prove the converse and direct parts of Theorem \ref{Thm:RT} in following two subsections, respectively. Theorem \ref{Thm:RT} only lower bounds the key rate. Clearly there is a tradeoff between the key rate and the coding rate. We propose an achievable region in Section \ref{Sec:Achiev}.

\subsubsection{Converse Part}\label{Sec:ZDParseCon}
For every lossless secure encoder--decoder pair that satisfies \eqref{eq:RT_Markov}, we lower bound the key rate as follows:
\begin{align}
 	\sum_{t=1}^n \bE l(K_t) &= \sum_{t=1}^n L(K_t) \nl
	&\geq \sum_{t=1}^n L(K_t|K^{t-1},Z^t)\label{eq:RT_Key0}\\
	&= \sum_{t=1}^n L(K_t,X_t|K^{t-1},Z^t)\label{eq:RT_Key1}\\
	&\geq \sum_{t=1}^n L(X_t|K^{t-1},Z^t)\label{eq:RT_Key2}\\
	&= \sum_{t=1}^n L(X_t|Z^t)\label{eq:RT_Key3}\\
	& =\sum _{t=1}^n L(X_t)\label{eq:RT_Key4}\\
    &=nL(X).
\end{align}
The first equality is true since the key bits are incompressible and therefore the Huffman length is the same as the number of key bits. \eqref{eq:RT_Key0} is true since conditioning reduces the Huffman length (the simple proof of this is omitted). \eqref{eq:RT_Key1} follows since $X_t$ is a function of $(K^t,Z^t)$ (the decoder's function) and therefore, given $(K^{t-1},Z^t)$, the code for $K_t$ also reveals $X_t$. \eqref{eq:RT_Key2} is true since with the same conditioning on $(K^{t-1},Z^t)$, the instantaneous code of $(K_t,X_t)$ cannot be shorter then the instantaneous code of $X_t$. \eqref{eq:RT_Key3} is due to \eqref{eq:RT_Markov} and finally, \eqref{eq:RT_Key4} is true since we consider a secure encoder. We therefore showed that $R_k\geq L(X)$.

\subsubsection{Direct Part}\label{Sec:RT_Direct}
We construct an encoder--decoder pair that are fully secure with $R_k = L(X)$. Let $l_{H-\max}$ denote the longest Huffman codeword of $X$. We know that $l_{H-\max}\leq |X|-1$. The encoder output will always be $l_{H-\max}$ bits long and will be built from two fields. The first field will be the Huffman codeword for the observed source symbol $X_t$. Denote its length by $l_H(X_t)$. This codeword is then XORed with $l_H(X_t)$ key bits. The second field will be composed of $l_{H-\max}-l_H(X_t)$ random bits (taken from the private source of randomness) that will pad the encrypted Huffman codeword to be of length $l_{H-\max}$. Regardless of the specific source output, Eve sees constant length codewords composed of random uniform bits. Therefore no information about the source is leaked by the encoder outputs. When Bob receives such a block, it starts XORing it with key bits until it detects a valid Huffman codeword. The rest of the bits are ignored. Obviously, the key rate which is needed is $L(X)$.

\subsubsection{Achievable Region} \label{Sec:Achiev}
\begin{figure}[htp]
\centering
\includegraphics[width=0.6\textwidth]{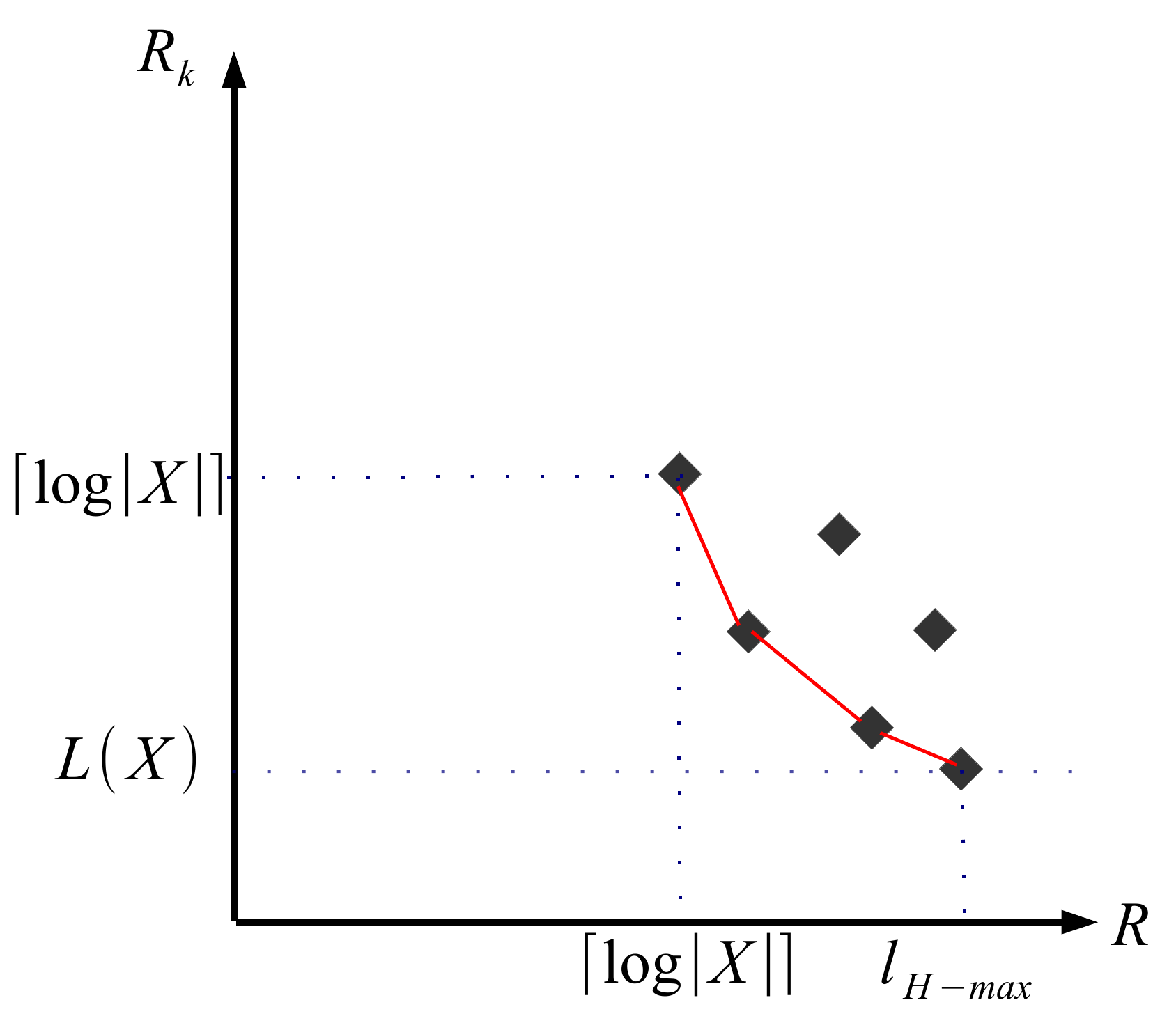}
\caption{Achievable region}\label{Fig:AchvRgn}
\end{figure}

The direct proof shown above, suggests a single point on the $R-R_k$ plane. Namely, the point where $R_k=L(X)$ (its minimal possible value), but the encoder rate is high and is equal to $l_{H-\max}$. However, there are many other achievable points which can be shown using the same idea as in the direct part of the proof by replacing the Huffman code by another instantaneous code. For every possible instantaneous code for the source $X$ (not necessarily optimal code for this source) we can repeat the arguments of the direct part of the proof by setting the rate (block length) to be the longest codeword in this code and the XORing only the bits which originate from this code while padding the remainder of the block with random bits. Any such code, which is a sub-optimal code for source coding, will give us an operating point in the $R-R_k$ plane. The extreme case is obtained by using the trivial source code which uses $\lceil \log |\calX| \rceil$ bits for all symbols and XORing all the bits with key bits. Since there is only a finite number of possible instantaneous codes (when not counting codes that artificially enlarge the description of shorter codes by adding bits) we will get a finite number of points in the $R-R_k$ plane. The lower convex envelope of these points is achievable by time-sharing. This is illustrated in Fig. \ref{Fig:AchvRgn}. We see an interesting phenomenon where optimal codes for source coding (e.g, Huffman code) will achieve high encoding rates (albeit low key rate) while sub-optimal codes will achieve low encoding rate (but higher key rate).

%%%%%%%%%%%%%%%%%%%%%%%%%%%%%%%%%%%%%%%%%%%%%%%%%%%%%%%%%%%%%%%%%%%%%
\subsection{Eve Has No Parsing Information}\label{Sec:ZD_NoParsing}
In this subsection, we relax our secrecy requirements and assume that Eve observes the whole transmission from Alice to Bob, but has no information on how to parse the bit-stream $B_n$ into $Z_1,\ldots, Z_n$. Note that we did not restrict the eavesdropper in any way since parsing information is generally not available and cannot be calculated from $B_n$. Moreover, this setting has a practical motivation, as discussed in the Introduction.

Note that although Eve has no parsing information, she still knows the length of the whole bit-stream, which can still leak information on the source. For example, suppose Alice uses a prefix code that has a 1 bit codeword length for, say, $X=a$ and sends $n$ symbols to Bob. Even if the bit-stream is encrypted with a one time pad (and therefore not parsable by Eve by our assumptions) but Eve sees that the total length of the bit-stream is exactly $n$ bits, she knows with certainty that all the symbols in the bit-stream are $a$. While such a system is not secure by Definition \ref{def_RTSecure}, the probability of the described event vanishes exponentially fast. 
In this section, we have a relaxed definition of secrecy which allows for encoders that leak information on the source, but with vanishing probability:

\begin{mydef}
When Eve has no parsing information, we say that the system is asymptotically secure if the following holds for every $t\leq n$ and every $x\in\calX$:
\begin{align}
 P(X_t=x|B_n) \xrightarrow[n\to\infty]{} P_X(X_t=x)~~ a.s.\label{eq:RT_NoParseSecCon}
\end{align}
\end{mydef}
This means that when the bit-stream is long enough, the eavesdropper does not learn from it anything about the source symbols with probability 1. Note that the encoder from Section \ref{Sec:RT_Direct} trivially satisfies this constraint since it was a constant block length encoder and the bits within the block where encrypted by a one--time pad. We will see that with the relaxed secrecy requirement we can reduce the rate of the encoder to be the same as the rate of the key. 
As in the previous subsection, where we dealt with encoders satisfying \eqref{eq:RT_Markov}, here, we will limit the discussion as well. We will treat only encoders that satisfy
\begin{align}
 	\lim_{n\to\infty}\max_{1\leq t\leq n(1-\epsilon)}\|P(x_t|B_n,k^{t-1})-P(x_t|B_n)\|_1 =0 \text{    } a.s. \label{eq:NoParseConstraint}
\end{align}
where $\| P-Q \|_1$ denotes the variational distance between the probability measures, $ \| P-Q \|_1 = \sum_x |P(x)-Q(x)|$. This constraint means that when $n$ is large, for all $t$ for which a gap remains to the end of the bit-stream, we practically have the Markov chain as in \eqref{eq:RT_Markov} with $Z^n$ replaced by $B_n$. Note that this requirement is less stringent than \eqref{eq:RT_Markov} and in fact any encoder which satisfies \eqref{eq:RT_Markov} and is secure with parsing information (i.e., satisfies definition \ref{def_RTSecure}) will satisfy \eqref{eq:NoParseConstraint}. Since $B_t$ can be parsed with $K^t$, the margin between $t$ and $n$ in \eqref{eq:NoParseConstraint} ensures that for any $t$ considered, there is a portion of $B_n$ that cannot be parsed, even if the eavesdropper acquired the previous keys. The discussion that followed the constraint \eqref{eq:RT_Markov} is valid here as well, namely, we only consider encoders that do not reuse old key bits to encrypt new messages. 

We have the following theorem:
\begin{theorem}\label{Thm:RT_NoParse}
	In the class of codes that satisfy  \eqref{eq:NoParseConstraint}, there exists a pair of asymptotically secure, zero-delay, encoder and decoder if and only if	
 	$R \geq L(X), R_k \geq L(X)$.
\end{theorem}
The fact that $R\geq L(X)$ is trivial since we deal with a zero-delay lossless encoder. However, unlike the case of Theorem \ref{Thm:RT}, here it can be achieved along with $R_k\geq L(X)$. Note that if instead of defining the secrecy constraint as in \eqref{eq:RT_NoParseSecCon}, we required that for every $n,t$, $P(X_t|B_n)=P(X_t)$ then a counterpart of Theorem 1 will hold here. However, the encoder will, as in the proof of the direct part of Theorem \ref{Thm:RT}, operate at constant rate.
We prove the converse and direct parts of Theorem \ref{Thm:RT_NoParse} in the following two subsections, respectively.

\subsubsection{Converse Part}
Following the arguments of Section \ref{Sec:ZDParseCon}, we have:
\begin{align}
 	\sum_{t=1}^n \bE l(K_t) &= \sum_{t=1}^n L(K_t) \nl
	&\geq \sum_{t=1}^n L(K_t|B_n,K^{t-1})\nl
	&= \sum_{t=1}^n L(K_t,X_t|B_n,K^{t-1})\nl
	&\geq \sum_{t=1}^n L(X_t|B_n,K^{t-1})\nl
	&\geq \sum_{t=1}^{n(1-\epsilon)} L(X_t|B_n,K^{t-1}).
\end{align}
Now, let us define $\calB_n=\cbr{\bb_n: \max_{1\leq t \leq n(1-\epsilon)}\|P(x_t|\bb_n,k^{t-1})-P(x_t|\bb_n)\|_1\leq \epsilon, ~~\forall k^{t-1}}$. 
It can be immediately seen that for $\bb_n\in\calB_n$, 
\begin{align}
	L(X_t|\bb_n,K^{t-1})\geq L(X_t|\bb_n)-\epsilon|\calX|l_{\max}. 
\end{align}
where $l_{\max}$ is the longest codeword for $X$, whose length can be bounded by $|\calX|$. 
Using the definition given in \eqref{eq:CondHuffman}, we continue as follows:
\begin{align}
	\sum_{t=1}^n \bE l(K_t) &\geq \sum_{t=1}^{n(1-\epsilon)} L(X_t|B_n,K^{t-1})\nl
	&\geq \sum_{t=1}^{n(1-\epsilon)} \sum_{\bb_n\in\calB_n}P(\bb_n)L(X_t|\bb_n,K^{t-1})\nl
	& \geq \sum_{t=1}^{n(1-\epsilon)} \sum_{\bb_n\in\calB_n}P(\bb_n)L(X_t|\bb_n) -n(1-\epsilon)\epsilon |X| l_{max}
\end{align}
We now define $\calC_n=\cbr{\bb_n: \| P(x_t|\bb_n)-P(x_t)\|_1\leq \epsilon}$ and continue as follows
\begin{align}
 	\sum_{t=1}^n \bE l(K_t)  &\geq \sum_{t=1}^{n(1-\epsilon)} \sum_{\bb_n\in\calB_n}P(\bb_n)L(X_t|\bb_n) -n(1-\epsilon)\epsilon |X| l_{max}\nl
	& \geq \sum_{t=1}^{n(1-\epsilon)} \sum_{\bb_n\in\calB_n\cap\calC_n}P(\bb_n)L(X_t|\bb_n) -n(1-\epsilon)\epsilon |X| l_{max}\nl
	& \geq \sum_{t=1}^{n(1-\epsilon)} \sum_{\bb_n\in\calB_n\cap\calC_n}P(\bb_n)L(X_t) -2n(1-\epsilon)\epsilon |X| l_{max}\nl
	&=n(1-\epsilon)P(B_n\cap C_n) L(X) -2n(1-\epsilon)\epsilon |X| l_{max}
\end{align}
Now since by \eqref{eq:RT_NoParseSecCon} and \eqref{eq:NoParseConstraint} we have that $P(B_n\cap C_n)\to 1$ for any $\epsilon>0$ we showed that
\begin{align}
 	R_k=\limsup_{n\to\infty}\frn\sum_{t=1}^n \bE l_{K_t} \geq L(X).
\end{align}

\subsubsection{Direct Part}
 
The direct part of the proof is achieved by separation. We show that the simplest memoryless encoder that encodes $X_t$ using a Huffman code and then XORs the resulting bits with a one time pad is optimal here. Therefore, both the coding rate and the key rate of this scheme are equal to $L(X)$. Note that with such a simple encoder, no prior knowledge of $n$ is needed and therefore such an encoder is suitable for real-time streaming applications. We need to show that \eqref{eq:RT_NoParseSecCon} holds. 

We outline the idea before giving the formal proof. The bits of $B_n$ are independent of $X_t$ since we encrypted them with a one-time pad. Therefore, only the total length of the bit-stream can leak information. Let $l(B_n)$ represent the number of bits in $B_n$. By the strong law of large numbers, $\frn l(B_n)\to L(X)~a.s$. But if $l(B_n)\approx nL(X)$ then the only thing that Eve learns is that $X^n$ is typical (or equivalently that the law of large number is working) and this, of course, is known in advance. There are events where Eve can indeed learn a lot from the length of $B_n$, but the probability of these events vanish as $n$ becomes large. 

We invoke martingale theory \cite{DurrettBook} for a formal proof. Note that $B_n$ can be written as $B_n = (l(B_n),V_n)$, where $V_n$ is uniform over $\cbr{1,2,\ldots,2^{l(B_n)}}$ and is the number which is represented by the bits of $B_n$ (in base 2). Given $l(B_n)$, $V_n$ is independent of $X^n$ since all bits are encrypted. Therefore, we have the following chain $X_t\leftrightarrow l(B_n)\leftrightarrow B_n$. Now, since $l(B_n)$ is a function of $B_n$, we have that $P(X_t|B_n)=P(X_t|B_n,l(B_n))=P(X_t|l(B_n))$. 

Let $Y_t = l_H(X_t)$ where $l_H(X_t)$ represents the Huffman codeword length associated with $X_t$. Clearly, $l(B_n) = \sum_{i=1}^n Y_i$. Also for $x\in\calX$ define the indicator $\calI_t(x) = \ind{X_t=x}$. Finally, define the filtration 
\begin{align}
	\calF_{-n} &= \sigma\cbr{l(B_n), l(B_{n+1}), \dots}
	%&=\sigma\cbr{l(B_n), Y_{n+1}, Y_{n+2}, \ldots} 
\end{align}
Since the $Y_i$-s are independent and since the events in $\calF_{-\infty}$ (which depend on $\cbr{Y_i}_{i=1}^{\infty}$) are invariant to a finite permutation of the indexes of the $Y_i-s$, we have by the Hewitt-Savage zero--one law \cite[Theorem 4.1.1, p. 162]{DurrettBook} that $\calF_{\infty}$,  where $\calF_{-\infty} = \cap_{n}\calF_{-n}$, is trivial. Namely, if $\calA\in\calF_{-\infty}$, $P(\calA)\in\cbr{0,1}$. 

For $n\geq t$, let $M_n = \bE\sbr{\calI_t(x) | \calF_{-n}}$. Note that $M_n$ is a bounded (reverse) martingale which converges almost surely to $M_{\infty}$ as $n\to\infty$. Since the source is memoryless and since $n\geq t$, we have that $\calI_t(x)$ is independent of $Y_{n+1},Y_{n+1,\ldots}$ and therefore given $l(B_n)$, $X_t$ is independent of $l(B_{n+1}), l(B_{n+2}), \ldots$. Thus we have that  $M_n = \bE\sbr{\calI_t(x) | l(B_n), l(B_{n+1}), \ldots}= P(X_t=x|l(B_n))$.  Also note that $M_n$ does not depend on $t$, since due to symmetry, $P(X_t=x|l(B_n))=P(X_1=x|l(B_n))$. This allows us to consider not only finite $t$, but also $t\leq n$ that grow with $n$ to infinity.

Finally, the fact that $\calF_{\infty}$ is trivial implies that $M_{\infty}$ is a constant, which is the expectation of the indicator, i.e., $P(X_t=x)$. We therefore showed that $P(X_t=x|l(B_n)) \to P(X_t=x) ~ a.s$.

To see that this encoder satisfies \eqref{eq:NoParseConstraint} note that 
\begin{align}
	P(X_t=x_t|B_n,k^{t-1})&=P(X_t=x_t|B_{t,n},x^{t-1})\nl
	&=P(X_t=x_t|B_{t,n})\nl
	&=P(X_t=x_t|l(B_n)-l(B_{t-1}))
\end{align}
where $B_{t,n}$ denotes the bit-stream starting from time $t$ until time $n$. The last equation is true since we have that $X_t\leftrightarrow l(B_n)-l(B_{t-1})\leftrightarrow B_{t,n}$
Now, for $t\leq (1-\epsilon)n$, the analysis we did above for $P(X_t=x_t|B_n)$ is valid here (with a time shift of $t$ ) since $l(B_n)-l(B_{t-1})$ grows as $n$ grows and therefore  \eqref{eq:NoParseConstraint} is satisfied.  

\section{Causal Rate-Distortion with Secrecy Constraints}\label{Sec:CausalRD}
In this section, we extend \cite{NeuhoffGilbert1982},\cite{TsachyNeri2005},\cite{Me-RTRD2013} to include secrecy constraints. Unlike the discussion of Section \ref{Sec:RealTimeSecrecy}, here we will allow lossy reconstruction and imperfect secrecy. 

As in any source coding work, we will be interested in the encoder rate $R$ and the minimal distortion, $D$ which is attained with this rate. With the addition of secrecy constraints, two more figures of merit will be added. The first is the uncertainty $h$, (measured by equivocation, to be  defined shortly) at the eavesdropper regarding the source. The second is the rate of a private key, $R_k$, shared by Alice and Bob with which a given uncertainty $h$, is achievable. Note that when $D>0$, Bob is also left with uncertainty regarding the original source sequence since he only knows it is contained in a $D$-ball around its reconstruction. Therefore, even with no attempt of encryption, Eve will ``suffer'' this same uncertainty. This implies that $(R,R_k,h,D)$ are tied together and the goal of this section is to find the region of attainable quadruples.   

\begin{figure}[htp]
\centering
\includegraphics[width=0.9\textwidth]{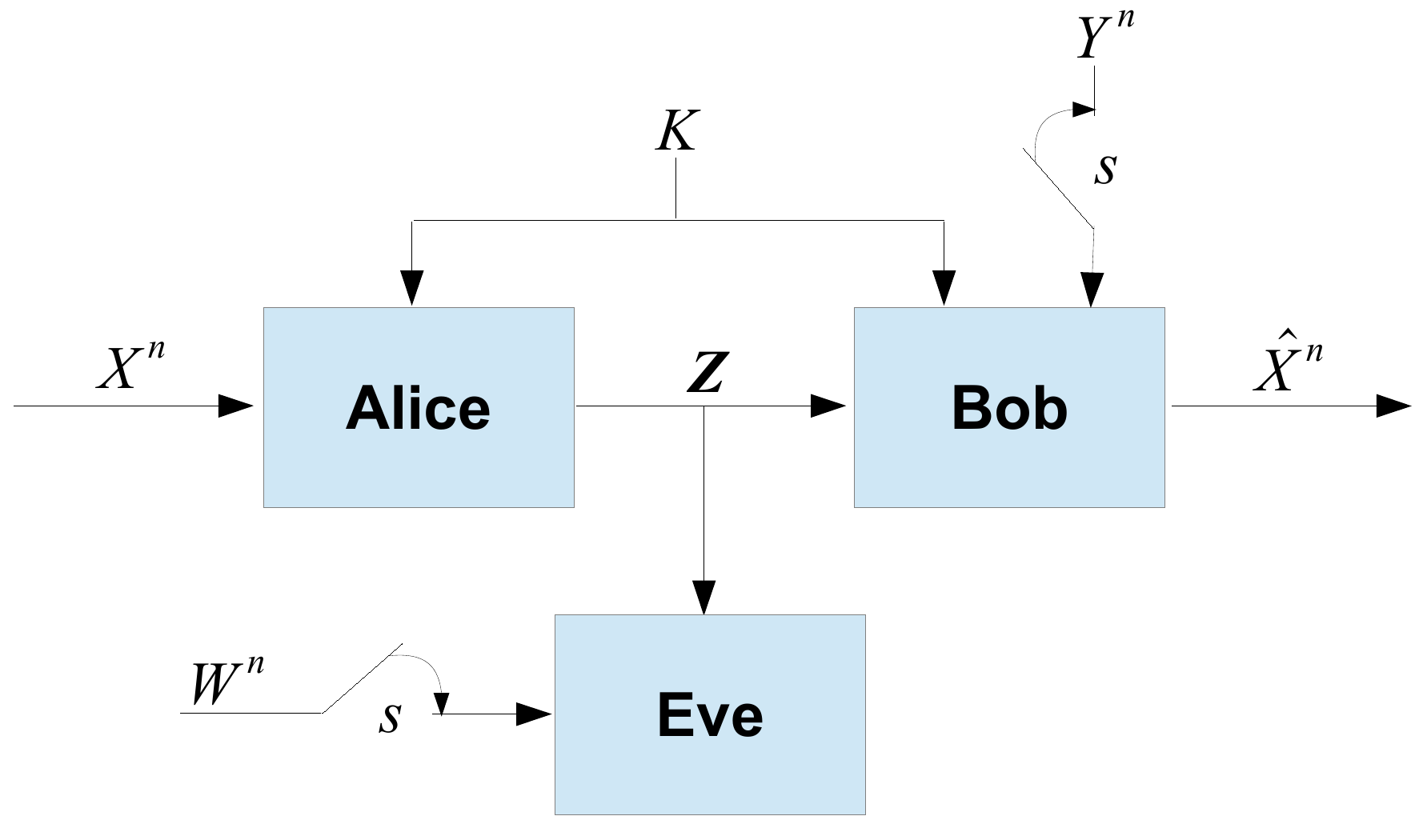}
\caption{Causal Model}\label{Fig:CausModel}
\end{figure}

Our system model is depicted in Fig. \ref{Fig:CausModel}.

We will deal with two settings, which differ by the position of the switches denoted by $S$ in Fig. \ref{Fig:CausModel}. Namely, the availability of SI at Bob and Eve. While the setting without SI is a special case of the setting which includes SI, these settings are different from a causal rate-distortion standpoint. Without SI, it was shown in \cite{NeuhoffGilbert1982} that the chain of encoder and decoder is equivalent to chain of a causal reproduction coder (to be formally defined shortly) followed by lossless compression. However, when SI is available to the decoder, this equivalence does not hold since the encoder cannot reproduce $\hX$ without the SI.
The next two subsections deal with the two settings, starting from the simpler case without SI. Formal definition of the settings will be given in the beginning of each subsection.

\subsection{Causal Secrecy Without SI} \label{Sec:CausalNoSI}
\subsubsection{Preliminaries}
The notation conventions we introduced in the beginning of Section \ref{Sec:ZDPrelim} for random variables, vectors etc., will be used here as well. 
Let $X^n$ be a sequence produced by a memoryless source. The alphabet of $X$, $\calX$, as well as all other alphabets in the sequel, is finite. The source sequence is given to Alice. In addition to the source, Alice has access to a secret key, $K$, uniformly distributed over $\cbr{1,2,\ldots,M_k}$, which is independent of $X^n$. Bob shares the same key. Alice uses the source and the key to create a binary representation $\bZ=\cbr{Z_k}_{k\geq 1}$ (we omit the dependence of $\bZ$ on $n$ to simplify notation). Bob receives $\bZ$ and with its shared key creates a reproduction $\hX^n$, where $\hX\in\hat{\calX}$ is the reproduction alphabet. As in \cite{NeuhoffGilbert1982}, the cascade of encoder and decoder will be referred as a \textit{reproduction coder}, i.e., the reproduction coder is a family of functions $\cbr{f_k}_{k\geq 1}$ such that $\hX_k=f_k(X^n,K)$. 

We say that a reproduction function is causal relative to the source if for all $t$:
\begin{align}
    \hX_t = f_t(X_{-\infty}^{\infty},K) = f_t(\tilde{X}_{-\infty}^{\infty},K) \text{ if } X_{-\infty}^{t}=\tilde{X}_{-\infty}^{t}.
\end{align}
%Let $\calF_c$ denote the class of all encoder-decoder pairs which induce a causal reproduction coder, relative to the source
We are given a distortion measure $d$, $d: \calX\times\hat{\calX}\to\mathbbm{R}^+$ where $\mathbbm{R}^+$ denotes the set of non-negative real numbers. Let $D_{\min}=\min_{x,\hx}d(x,\hx)$. When a source code with a given induced reproduction coder $\cbr{f_k}$ is applied to the source $X^n$, the average distortion is defined by
\begin{align}
	D(\cbr{f_k})\eqd \limsup_{n\to\infty}\frn \bE \sum_{t=1}^n d(X_t,\hX_t).  \label{eq:DefDist}
\end{align}
Let $l(\bZ)$  denote the number of bits in the bitstream $\bZ$. The average rate of the encoder is defined by
\begin{align}
	R = \limsup_{n\to\infty} \frn\bE l(\bZ). \label{eq:DefRate}
\end{align}

In our model, an eavesdropper, Eve, intercepts the bitstream $\bZ$. We follow the common assumptions that Eve is familiar with the encoding and decoding functions, coding techniques, etc., which are employed by Alice and Bob, but is unaware of the actual realizations of the source and the key. The uncertainty regarding the source sequence $\bX$ at the eavesdropper after intercepting $\bZ$ is measured by the per-letter equivocation, which we denote by $h$. Namely, $h=\liminf_{n\to\infty}\frn H(X^n|\bZ)$.

Unlike \cite{NeuhoffGilbert1982}, \cite{TsachyNeri2005}, where the bit representation which is passed from the encoder to the decoder, $\bZ$, was only an intermediate step between a lossless encoder and a lossless decoder, here, $\bZ$ is important as it should leave Eve as oblivious as possible of the source sequence. However, as in \cite{NeuhoffGilbert1982}, \cite{TsachyNeri2005}, applying a lossless code on $\bZ$ between the encoder and decoder can only improve the coding rate and will not affect the other figures of merit.

Let $\calR_{NO-SI}$ denote the set of positive quadruples $(R,R_k,D,h)$ such that for every $\epsilon>0$ and sufficiently large $n$, there exists an encoder and a decoder, inducing a causal reproduction coder and satisfying:
\begin{align}
     \frn H(\bZ) &\leq R+\epsilon, \label{eq:R}\\
    \frn H(K) &\leq R_k +\epsilon, \label{eq:R_k}\\
    \frn\sum_{t=1}^n\bE d(X_t,\hX_t)&\leq D +\epsilon, \\
    \frn H(X^n | \bZ)&\geq h -\epsilon. \label{eq:Equivo1}
\end{align}
Our goal, in this section, is to find the region of quadruples, which are achievable with a causal reproduction coder. To this end, we will need previous results on causal rate-distortion results \cite{NeuhoffGilbert1982}, which we briefly describe below.

\begin{figure}[htp]
\begin{subfigure}{\textwidth}
\centering
\includegraphics[width=0.8\textwidth]{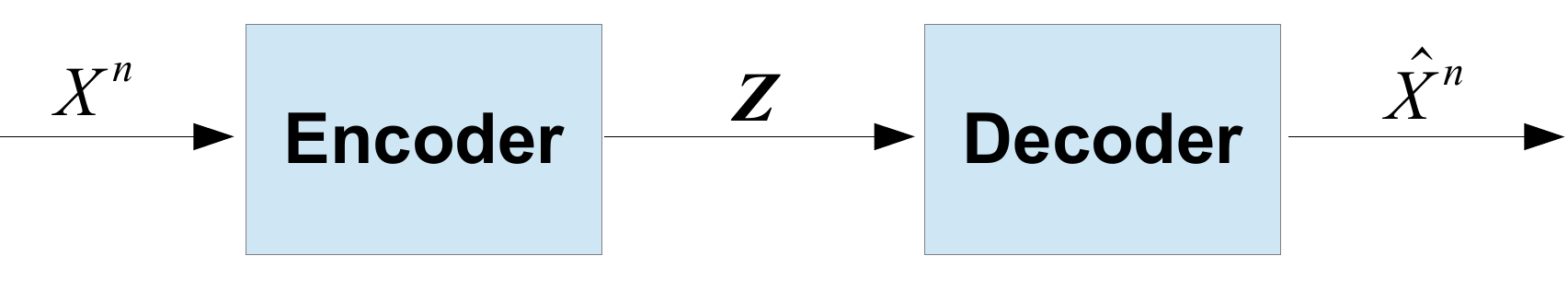} \label{Img:Orig}
\caption{Original source coding model}
\end{subfigure}

\begin{subfigure}{\textwidth}
\centering
\includegraphics[width=0.8\textwidth]{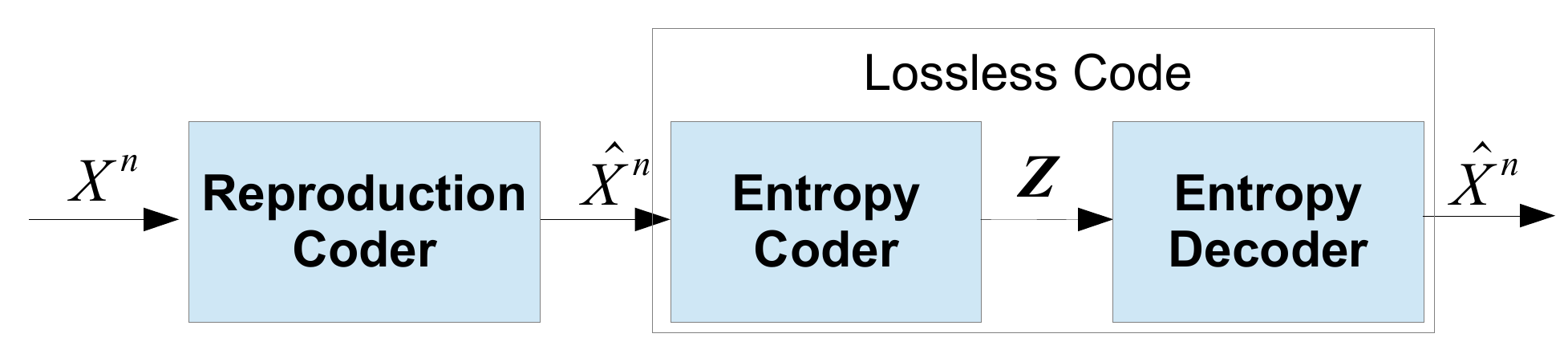} \label{Img:Equiv}
\caption{Equivalent source coding model}
\end{subfigure}
\caption{Causal source coding model}\label{Fig:NHModels}
\end{figure}

In \cite{NeuhoffGilbert1982}, the same model as described above was considered without the secrecy constraints. The goal of \cite{NeuhoffGilbert1982} was to find the tradeoff between $R$ and $D$ under the constraint that the reproduction coder is causal. Towards this goal, the equivalence of the two models in Fig. \ref{Fig:NHModels} was proved. Namely, the rate of a source code with a given reproduction coder can be only improved if a lossless source code will be applied to its output. This implied that it is enough to consider only systems that first, generate the reproduction processes and then losslessly encode it as is Fig. \ref{Fig:NHModels}b. Indeed, this separation is the basis for many practical coding systems. 

Let $\calR_c(D)$ denote the minimal achievable rate over all causal reproduction coders $\cbr{f_k}$ with $d(\cbr{f_k})\leq D$. Also, let
\begin{align}
 	r_c(D) = \min_{f:\boldmath{E} d(X,f(X)\leq D} H(f(X))
\end{align}
and finally, let $\underline{r}_c(D)$ denote the lower convex envelope of $r_c(D)$ 
The following theorem is the main result of \cite{NeuhoffGilbert1982}:
\begin{theorem}(\cite{NeuhoffGilbert1982}, Theorem 3) \label{Thm:NGThm}
\begin{align}
	\calR_c(D)=\underline{r}_c(D).
\end{align}
\end{theorem}
It was shown in \cite{NeuhoffGilbert1982} that $\underline{r}_c(D) $ is achieved by time-sharing no more than two scalar reproduction coders.

\subsubsection{Characterizing $\calR_{NO-SI}$}

Although a scheme that first uses the optimal encoder of \cite{NeuhoffGilbert1982} to create the same bit-stream and then XOR the resulting bits with key bits  is obviously possible, it is not immediately clear that such a separation is optimal. The reason for this is that one needs to first rule out the possibility that using key bits during the quantization and using entropy coding with the key as SI at both sides might improve performance. The following theorem, which is the main result of this subsection, shows that the separation scheme is indeed optimal.
\begin{theorem}\label{Thm:CausalNoSI}
	 $(R,R_k,D,h)\in\calR_{NO-SI}$ if and only if the following inequalities hold:	 
\begin{align}
	R &\geq  \underline{r}_c(D) \nl
	h &\leq H(X) \nl
	R_k &\geq h-H(X)+\underline{r}_c(D) 
\end{align}
\end{theorem}

It is evident from Theorem \ref{Thm:CausalNoSI} that the direct part is achieved by the separation scheme proposed above. Theorem \ref{Thm:CausalNoSI} is a special case of the more general theorem, which includes SI, we prove in the next subsection (Theorem \ref{Thm:CausalSI}) and therefore no proof is given here\footnote{The achieving scheme of Theorem \ref{Thm:CausalNoSI}  will use the coding scheme of \cite{NeuhoffGilbert1982} instead of the coding scheme of \cite{Me-RTRD2013} which is used to prove Theorem  \ref{Thm:CausalSI}, but the ideas are essentially the same.}.

\subsection{Causal Secrecy with Side Information}
%%%%%%%%%%%%%%%%%%%%%%%%%%%%%%%%%%%%%%%%%%%%%%%%%%%%%%%%%%%%%%%%%
In this section, we extend the setting of Section \ref{Sec:CausalNoSI} to include SI at Bob and Eve. As in the previous section, we start by describing the model and mentioning related causal rate-distortion results before giving our results. 

In this section, we assume a memoryless source which emits sequences of random variables $(X^n,Y^n,W^n)$. As before, $X^n$ is observed by Alice. Bob and Eve observe $Y^n$, $W^n$ respectively.
The sequences $(X^n, Y^n, W^n)$ are distributed according to a joint distribution $$P(x^n,y^n,w^n) = \prod_{t=1}^n P(x_t)P(y_t|x_t)P(w_t|y_t).$$ Namely, we assume a degraded SI structure where Bob's SI is better than Eve's SI. Although we do not deal with other possible SI structures, we will discuss such extensions in the sequel.
%Unlike \cite{Prab-Ramch1997}, \cite{VillardPaint2011}, we will treat in this section only the case of degraded SI. This model covers the scenarios where no SI is available or is available only to Bob as special cases.
As in the model without SI, Alice and Bob have access to a shared secret key denoted by $K$, $K \in \{0,1,2\ldots,M_k\}$ which is independent of the source.
Alice sends a binary representation $\bZ=\cbr{Z_k}_{k\geq1}$ which is used by the decoder along with the SI and key to create $\hX^n$. We call the cascade of encoder and decoder a \textit{reproduction coder}, namely a family of functions $\cbr{f_k}_{k\geq 1}$ such that $\hX_k=f_k(K, X^n,Y^n)$. A reproduction coder is said to be causal with respect to the source and SI if for all $t$:
\begin{align}
	f_t(k,x^n,y^n) = f_t(k,u^t,v^t), \text{   if   } x^t=u^t, y^t=v^t 
\end{align}
The distortion constraint \eqref{eq:DefDist} and the definition of rate \eqref{eq:DefRate} remain the same as in the setting without SI. 
When SI is available to the decoder, the cascade of encoder and decoder cannot be a recast into a cascade of reproduction coder followed by lossless coding as in Fig. \ref{Fig:NHModels} since the encoder has no access to the SI the decoder uses improve its reproduction and therefore cannot calculate $\hX_t$. A \textit{causal reproduction coder} in our setting is composed of a family of causal encoding functions which calculate messages, $S_t$, which are causal functions of the source symbols, namely $S_t=e_t(K, X^t)$ and causal decoding functions which use the encoder messages along with the SI and key to create the reproduction, namely, $\hX_t=g_t(K, S^t,Y^t)$. Note that this representation stems directly from the causality restriction and every system that induces a causal reproduction coder can be written in this way\footnote{The causality restriction does not actually force that $S_t$ will be a function of $X^t$ as long as when reproducing $\hX_t$, only $S_i$'s which are functions of $(X_1,X_2,\ldots,X_t)$ are used. However, such an ``out of order'' system has an equivalent system which is obtained by reordering the indexes of the messages so that $S_i$ is a function of $X^i$. The performance of both systems will be equivalent since the reordering does not affect the neither the entropy coding of $S^n$ nor the calculation of $\hX_t$}. 
As in \cite{NeuhoffGilbert1982}, although forcing causal encoding and decoding functions, this model allows for arbitrary delays which can be introduced when transmitting the encoder's output, $S^n$, to the decoder. Namely, $\bZ$ is not necessarily causal in $S^n$. We will allow events where the decoder fails to decode the bit-stream $\bZ$ to produce the encoder messages, $S^n$. In such an event, we have no restriction on the dependence of the output on the SI (and therefore on the source). However, we restrict that such error events will happen with small probability. Namely, we require that for every $\epsilon>0$, $P(S^n\neq g(K,Y^n,\bZ))<\epsilon$ for large enough $n$ and some function $g$.

%When SI is available to the decoder, Slepian-Wolf coding is possible. However, unlike the entropy coding which was used when no SI was available and could be done with zero error probability, this is not possible with Slepian-Wolf coding.  
%When the Slepian wolf $M^n=g(K,Y^n,\bZ)$, with high probability and a failure event is allowed. Obviously, as long as the probability of this failure event remains negligible, this does not affect the average distortion. 

The secrecy of the system is measured by the uncertainty of Eve with regard to the source sequence, measured by the normalized equivocation $\frn H(X^n|W^n,\bZ)$. 

Let $\calR_{SI}$ denote the set of positive quadruples $(R,R_k,D,h)$ such that for every $\epsilon>0$ and sufficiently large $n$, there exists an encoder and a decoder inducing a causal reproduction coder satisfying:
\begin{align}
     \frn H(\bZ) &\leq R+\epsilon, \label{eq:R}\\
    \frn H(K) &\leq R_k +\epsilon,\label{eq:R_k}\\
    \frn\sum_{t=1}^n\bE d(X_t,\hX_t)&\leq D +\epsilon,\\
    \frn H(X^n|W^n,\bZ)&\geq h -\epsilon. \label{eq:Equivo}
\end{align}
Our goal is to characterize this region.

In the context of causal source coding, such a model was considered in \cite{TsachyNeri2005}, \cite{Me-RTRD2013}. In \cite{TsachyNeri2005} the model was restricted to common reconstruction between the encoder and decoder, meaning that both parties agree on the reconstruction. This restriction prevents the decoder from using the SI when reconstructing the source and the SI can be used only for lossless transmission of the reconstruction which is calculated at the encoder. In this case, a cascade of a reproduction coder followed by lossless entropy coding which uses SI is valid. The full treatment of SI for this scenario was recently given in \cite{Me-RTRD2013}. 

Let $\calR_c^{SI}(D)$ denote the minimal achievable rate over all causal reproduction coders with access to SI, $\cbr{f_k}$, such that $D\cbr{f_k}\leq D$. Also let
\begin{align}
	r_c^{SI}(D) = \min_{f,g} H(f(X)|Y) \label{eq:R_SIofD_DEF}
\end{align}
where the minimum is over all functions $f: \calX\to\calS$ and $g:\calS\times\calY\to\hat{\calX}$ such that $\bE d(X, g(f(X),Y))\leq D$. The alphabet $\calS$ is a finite alphabet whose size is part of the optimization process ($|\calS| \leq |\calX|$). Finally, let $\underline{r}_c^{SI}(D)$ be the lower convex envelope of $r_c^{SI}(D)$.

The following theorem is proved in \cite{Me-RTRD2013}:
\begin{theorem}(\cite{Me-RTRD2013}, Theorem 4) \label{Thm:PrevRDSI}
\begin{align}
	\calR_c^{SI}(D) = \underline{r}_c^{SI}(D) 
\end{align}
\end{theorem}
It was shown that $\underline{r}_c^{SI}(D)$ is achieved by time-sharing at most two sets of scalar encoders ($f$) and decoders ($g$). Moreover, SI lookahead was shown to be useless in the causal setting. 

\subsection{Characterizing $\calR_{SI}$}

We have the following theorem.

\begin{theorem}\label{Thm:CausalSI}
    $(R,R_k,D,h)\in\calR_{SI}$ if and only if
    \begin{align}
    	R&\geq \underline{r}_c^{SI}(D),\nt\\
         h&\leq H(X|W), \nl
         R_k&\geq h - H(X|W) +  \underline{r}_c^{SI}(D). \label{eq:Theorem}
    \end{align}
    If $h - H(X|W) +  \underline{r}_c^{SI}(D)\leq 0$, no encryption is needed.
\end{theorem}

\noindent \textit{Remark:} The above theorem pertains to degraded SI structure. It was shown in \cite{Prab-Ramch1997} that for lossless secure compression with SI, Slepian-Wolf coding is optimal when the SI is degraded, but not optimal otherwise. For a general SI structure, a simple scheme will apply memoryless quantization (resulting in a new memoryless source) and then apply the scheme of \cite{Prab-Ramch1997}. The output of the scheme of \cite{Prab-Ramch1997} can be further encrypted with key bits as needed to achieve the desired $h$. Such a scheme will not violate the causal restriction. Although schemes that first apply memoryless quantization and then losslessly compress the output are optimal in the context of all known causal source coding works, it is not clear that this is the case here. The challenge in the converse part is that when applying a causal encoder (but not memoryless), the resulting process is not necessarily memoryless. We were unsuccessful in proving the converse for a general SI structure. 
 
We prove the converse and direct parts of Theorem \ref{Thm:CausalSI} in the following two subsections, respectively.

\subsubsection{Converse Part}
We now proceed to prove the converse part, starting with lower bounding the encoding rate. We assume a given encoder and decoder pair which form a causal reproduction coder with $(R,R_k,D,h)\in\calR_{SI}$. By the definition of our model we have by Fano's inequality  (\cite{cover}) that $H(S^n|K,Y^n,\bZ)\leq n\epsilon$.

For $n$ large enough and every encoder and decoder pair that induce a causal reproduction coder and satisfy \eqref{eq:R}-\eqref{eq:Equivo} the following chain of inequalities hold:
\allowdisplaybreaks
\begin{align}
	nR &\geq H(\bZ) \nl
	&\geq H(\bZ|K,Y^n) - H(\bZ|K,S^n,Y^n)\nl
	&=I(S^n;\bZ|K,Y^n)\nl
	&=H(S^n|K,Y^n)-H(S^n|K,Y^n,\bZ)\nl
	&\geq H(S^n|K,Y^n)-n\epsilon\label{eq:SIConverse2}\\
	&=\sum_{t=1}^n H(S_t|K,S^{t-1},Y^n) -n\epsilon\nl
	&\geq \sum_{t=1}^n H(S_t|K,S^{t-1},X^{t-1},Y^n) -n\epsilon\nl
	&=\sum_{t=1}^n H(S_t|K,X^{t-1},Y^n)-n\epsilon\label{eq:SIConverse22}\\
	&=\sum_{t=1}^n H(e_t(K,X^{t-1},X_t)|K,X^{t-1},Y^n)-n\epsilon\nl
	&=\sum_{t=1}^n \int H(e_t(X_t,k,x^{t-1})|k,x^{t-1},y^{t-1},Y_t,Y^{n}_{t+1})d\mu(k,x^{t-1},y^{t-1})-n\epsilon \label{eq:SIConverseMu}\\
	&=\sum_{t=1}^n \int H(e_t(X_t,k,x^{t-1})|Y_t)d\mu(k,x^{t-1},y^{t-1})-n\epsilon
\end{align}
where \eqref{eq:SIConverse2} follows from Fano's inequality and in \eqref{eq:SIConverse22} we used the fact that $S^{t-1}$ is a function of $(K,X^{t-1})$. In \eqref{eq:SIConverseMu}, $\mu(\cdot)$ denotes the joint probability mass function of its arguments. In the last line, we used the independence of $X_t$ from the key and SI at time other then $t$. Now, $e_t(X_t,k,x^{t-1})$ can be seen as a specific function, $f$, in the definition of $r_c^{SI}(D)$ \eqref{eq:R_SIofD_DEF}. Also, with $(k,x^{t-1},y^{t-1})$ fixed, so is $s^{t-1}$ and the decoding function\\
$\hX_t=g_t(k,s^{t-1},S_t,y^{t-1},Y_t)$ can be seen as a specific choice of $g(\cdot,\cdot)$ in  \eqref{eq:R_SIofD_DEF}. With this observation, we continue as follows
\begin{align}
 	&n(R+\epsilon) \geq \sum_{t=1}^n \int H(e_t(X_t,k,x^{t-1})|Y_t)d\mu(k,x^{t-1},y^{t-1})\nl
	&\geq \sum_{t=1}^n\int r_{c}^{SI}(\bE [d(X_t, g_t(e_t(k,x^{t-1},X_t),s^{t-1},y^{t-1},Y_t))|k,x^{t-1},y^{t-1}])\times\nl
	&~~~~~~d\mu(k,x^{t-1},y^{t-1})\label{eq:RTConv1}\\
	&\geq \sum_{t=1}^n\int \underline{r}_{c}^{SI}(\bE [d(X_t, g_t(e_t(k,x^{t-1},X_t),s^{t-1},y^{t-1},Y_t))|k,x^{t-1},y^{t-1}])\times\nl
	&~~~~~~d\mu(k,x^{t-1},y^{t-1})\label{eq:RTConv2}\\
	&\geq \sum_{t=1}^n \underline{r}_{c}^{SI}\Big(\int\bE [d(X_t, h_t(e_t(k,x^{t-1},X_t),s^{t-1},y^{t-1},Y_t))|k,x^{t-1},y^{t-1}])\times\nl
	&~~~~~~d\mu(k,x^{t-1},y^{t-1}\Big)\label{eq:RTConv3}\\
	&\geq \sum_{t=1}^n \underline{r}_{c}^{SI}(\bE [d(X_t, h_t(e_t(K,X^t),S^{t-1},Y^{t}))])\nl
	&= \sum_{t=1}^n  \underline{r}_{c}^{SI}\rbr{ \bE \sbr{d(X_t, \hX_t)}}\nl
	&\geq n \underline{r}_{c}^{SI}\rbr{ \frn\sum_{t=1}^n\bE \sbr{d(X_t, \hX_t)}}\label{eq:RTConv4}\\
	&\geq n \underline{r}_{c}^{SI}\rbr{D},\label{eq:RTConv5}
\end{align}
where \eqref{eq:RTConv1} follows from the definition of $r_{c}^{SI}(D)$ and the discussion preceding the last equation block, \eqref{eq:RTConv2} follows from the definition of  $\underline{r}_{c}^{SI}(D)$,  \eqref{eq:RTConv3} and \eqref{eq:RTConv4} follow from the convexity of $\underline{r}_{c}^{SI}(D)$. Finally, \eqref{eq:RTConv5} follows from the fact that $\underline{r}_{c}^{SI}\rbr{D}$ is non-increasing in $D$.

%Now, let $S_t=(X^{t-1},K)$. We will use $/mu$ as a generic notation for a probability measure, with arguments revealing the random variable(s) to which it relate.
%Continuing from the last equation we have:
%\begin{align}
%	n(R +\epsilon) &\geq \sum_{t=1}^n H(f_t(X^{t-1},K,X_t)|K,X^{t-1},Y^n)\nl
%	&= \sum_{t=1}^n \int H(f_t(u,X_t)|S_t=u,Y^n=y^n)d\mu(u,y^n)\nl
%	&= \sum_{t=1}^n \int H(f_t(u,X_t)|Y_t=y_t)d\mu(u,y^n)\label{eq:SIConverse3}\\
%	&= \sum_{t=1}^n \int \sbr{\int H(f_{t,u}(X_t)|Y_t=y_t)d\mu(y_t)}d\mu(u)\nl
%	&= \sum_{t=1}^n \int \sbr{H(f_{t,u}(X_t)|Y_t)}d\mu(u)\nl
%	&\geq \sum_{t=1}^n \int \overline{r_{x|y}}(d(f_{t,u}))d\mu(u) \nl
%	&\geq \sum_{t=1}^n  \overline{r_{x|y}}\rbr{\int d(f_{t,u})d\mu(u)} \nl
%	&= \sum_{t=1}^n  \overline{r_{x|y}}(d(f_{t}) )\nl
%	&\geq n\overline{r_{x|y}}\rbr{\frn\sum_{t=1}^n d(f_{t})} \nl
%	&\geq n\overline{r_{x|y}}(D)
%\end{align}
%where \eqref{eq:SIConverse3} follows since $X_t$ is independent of $S_t$.
The key rate can be lower bounded as follows:
\begin{align}
&nR_k = H(K)\nl
&\geq H(K|\bZ,W^n)\nl
&=I(X^n;K|W^n,\bZ) + H(K|X^n,W^n,\bZ)\nl
&=H(X^n|W^n,\bZ)-H(X^n|K,W^n,\bZ)+H(K|X^n,W^n,\bZ)\nl
&\geq nh-H(X^n|K,W^n,\bZ)+H(K|X^n,W^n,\bZ)\nl
&\geq nh-H(X^n|K,W^n,\bZ)\label{eq:SIConverseStart}
\end{align}
where the line preceding the last is true due to \eqref{eq:Equivo}. We continue by focusing on $H(X^n|K,W^n,\bZ)$:
\begin{align}
&H(X^n|K,W^n,\bZ) \nl
%&= nh-H(X^n|K,W^n,\bZ)+H(X^n|K,W^n,Y^n,\bZ)-H(X^n|K,W^n,Y^n,\bZ)\nl
&= I(X^n;Y^n|K,W^n,\bZ)+H(X^n|K,Y^n,W^n,\bZ)\nl%\label{eq:SIConverse1}\\
%&= nh-H(Y^n|K,W^n,\bZ)+H(Y^n|K,X^n,W^n,\bZ)-H(X^n|K,Y^n,\bZ)\nl
&\leq H(Y^n|W^n)-H(Y^n|K,X^n,W^n,\bZ)+H(X^n|K,Y^n,\bZ)\label{eq:SIConverse04}\\
&= H(Y^n|W^n)-H(Y^n|X^n,W^n)+H(X^n|K,Y^n,\bZ)\label{eq:SIConverse4}\\
&= I(X^n;Y^n|W^n)+H(X^n|K,Y^n,S^n,\bZ) \nl
&~~~~~~~+ I(X^n;S^n|K,Y^n,\bZ)\nl
&= I(X^n;Y^n|W^n)+H(X^n|K,Y^n,S^n,\bZ) \nl
&~~~~~~~+ H(S^n|K,Y^n,\bZ) - H(S^n|K,X^n,Y^n,\bZ)\nl
&\leq nI(X;Y|W)+H(X^n|K,Y^n,S^n,\bZ) +n\epsilon \label{eq:SIConverse7}\\
%&= n(H(X|W)-H(X|Y)) +H(X^n|K,Y^n,\hX^n,\bZ)\nl
&\leq n(H(X|W)-H(X|Y)) +H(X^n|K,Y^n,S^n) \label{eq:SIConverse6}
\end{align}
where in \eqref{eq:SIConverse04} we used the degraded structure of the source. Eq. \eqref{eq:SIConverse4} is true since $\bZ$ is a function of $(K,X^n)$ and $K$ is independent of the source. Eq. \eqref{eq:SIConverse7} is true by Fano's inequality and the fact that $S^n$ is a function of $(K,X^n)$. Focusing on the last term of \eqref{eq:SIConverse6} we have
\begin{align}
%&= n(H(X|W)-H(X|Y)) +H(X^n|K,Y^n,\hX^n)-H(X^n|K,Y^n)-H(X^n|K,Y^n)\nl
&H(X^n|K,Y^n,S^n) = H(X^n|K,Y^n) -I(X^n;S^n|K,Y^n)\nl
&= nH(X|Y)-I(X^n;S^n|K,Y^n)\nl
&= nH(X|Y)-H(S^n|K,Y^n)+H(S^n|K,X^n,Y^n)\nl
&=nH(X|Y)-H(S^n|K,Y^n)  \label{eq:SIConverse5}\\
&\leq nH(X|Y)-nr_c^{SI}(D) +n\epsilon \label{eq:SIConverseEnd}
\end{align}
where \eqref{eq:SIConverse5} is true since $W^n$ is a function of $K,X^n$. Finally, the last line follows from \eqref{eq:SIConverse2}.
Combining \eqref{eq:SIConverseEnd} with \eqref{eq:SIConverse6} into \eqref{eq:SIConverseStart}, and using the arbitrariness of $\epsilon$, we showed that $R_k \geq h - H(X|W) +  \underline{r}_c^{SI}(D)$.

\subsubsection{Direct Part}
It is seen from Theorem \ref{Thm:CausalSI}, that separation holds in this case. The direct part of this proof is therefore straightforward: First, we apply the achieving scheme which achieves Theorem \ref{Thm:PrevRDSI} and was presented in \cite{Me-RTRD2013}. 
Namely, we find at most two encoding functions, $f_1,f_2$, along with the corresponding two decoding functions that use the SI, $g_1,g_2$, which achieve average distortion not greater than $D+\epsilon$. We apply the encoding functions to the source sequence with the appropriate time sharing to create the encoder messages, $S^n$. A Slepian-Wolf code is then applied on $S^n$. Let the resulting binary representation of the Slepian-Wolf codeword be denoted by $\bB$. By construction, we have that $H(\bB)\leq \underline{r}_c^{SI}(D)+\epsilon$ for $n$ large enough. We now XOR the first $n[h - H(X|W) +  \underline{r}_c^{SI}(D)]$ bits of $\bB$ with a one time pad $K$ of this length, creating the binary sequence $\bZ$ which is given to the decoder. We have
\begin{align}
	\frn H(K)&=h - H(X|W) +  \underline{r}_c^{SI}(D) \label{eq:directSI2} \\
	\frn H(\bZ) &\leq \underline{r}_c^{SI}(D)+\epsilon. \label{eq:directSI1} 
\end{align}
At the decoder, the Slepian-Wolf code is decoded (with failure probability smaller than $\epsilon$ for large enough $n$) and the decoding functions $g_1, g_2$ are applied on $S^n$ and the SI to create the reproduction. We need to show that the equivocation constraint is indeed satisfied with this scheme: 
\begin{align}
	H(X^n|W^n,\bZ) &= H(X^n|W^n) - I(X^n,\bZ|W^n)\nl
	&= nH(X|W) -H(\bZ|W^n) +H(\bZ|X^n,W^n)\nl
	&\geq nH(X|W) -H(\bZ)+H(\bZ|\bB,X^n,W^n)\nl
	&= nH(X|W )-H(\bZ)+H(\bB\oplus K|\bB)\label{eq:SIDirectMarkov}\\
	&= nH(X|W )-H(\bZ)+H(K)\nl
	&\geq n(h-\epsilon) \label{eq:directSI2}
\end{align}
where in \eqref{eq:SIDirectMarkov} we used the fact that by construction we have the following chain $W^n\to X^n\to\bB\to\bZ$. In \eqref{eq:directSI2}, we used \eqref{eq:directSI1} and \eqref{eq:directSI2}.
Note that when the decoding of the Slepian-Wolf code fails, the resulting (arbitrary) reconstruction sequence depends non-causally on the SI, thus breaking the causal reproduction coder structure. However, the probability of such event can be made negligible when  $n$ becomes large.

%%%%%%%%%%%%%%%%%%%%%%%%%%%%%%%%%%%%%%%%%%%%%%%%%%%%%%

\section{Conclusion}\label{Sec:Conclusions}

We investigated the intersection between causal and zero-delay source coding and information theoretic secrecy. It was shown that simple separation is optimal in the causal setting when we considered encoder and decoder pairs that are causal only with respect to the source. An interesting extension will be to investigate the setting where the use of the key is also restricted to be causal, for example when the key is streamed along with the source and should not be be stored. This will force the quantizer to encrypt the resulting sequence before the lossless code (see \cite{JohnsonIshwarSP2004} for example). In the zero-delay setting we considered only perfect secrecy. An interesting and algorithmically challenging research direction is to investigate imperfect secrecy in the zero-delay setting. Moreover, it was mentioned that the extension of our zero-delay results to the case when the same SI is available to Bob and Eve is straightforward. This will continue to hold when even if Bob and Eve have different versions of SI but $P(w,y) >0$ for all $w,y$, where $W$ is Eve's SI and $Y$ is Bob's SI. However, when imperfect secrecy is considered, it is not clear how different SI affects the equivocation at Eve. Such a setting is another direction for future research.
\pagebreak
\appendix
\noindent\huge{Appendix}

\normalsize

\subsection*{Proof of Markov chain}\label{App:MArkov}
We show that any secure encoder that satisfies $Z_t = f(K_t,V_t,X^t, Z^{t-1})$ satisfies the Markov chain.
To see why this is true observe that
\begin{align}
 	P(k^{t-1},x_t,z^t) = P(z^t)P(x_t|z^t)P(k^{t-1}|x_t,z^{t}) \label{eq:RT_ProbAll}
\end{align}

Focusing on $P(k^{t-1}|x_t, z^t)$ we have
\begin{align}
	P(k^{t-1}|x_t, z^t) &= \frac{P(k^{t-1},x_t,z^{t-1})P(z_t|k^{t-1},x_t,z^{t-1})} {P(x_t,z^t)} \nl
	&=  \frac{P(k^{t-1},x_t,z^{t-1})P(z_t|k^{t-1},x^t,z^{t-1})} {P(x_t,z^t)} \label{eq:app2}\\
	&=  \frac{P(k^{t-1},z^{t-1})P(x_t)P(z_t|x^t,z^{t-1})} {P(x_t,z^t)} \label{eq:app3}\\
	&=  \frac{P(k^{t-1},z^{t-1})P(x_t)P(z_t|z^{t-1})} {P(x_t)P(z^t|x_t)} \label{eq:app4}\\
	&=  \frac{P(k^{t-1},z^{t-1})P(z_t|z^{t-1})} {P(z^{t-1})P(z_t|z^{t-1})}\nl
	&=P(k^{t-1}|z^{t-1})
\end{align}
where in \eqref{eq:app2} we note that $x^{t-1}$ can be computed from $z^{t-1},k^{t-1}$. In \eqref{eq:app3} we used the independence of $(k_t,v_t)$ from $k^{t-1}$ and the fact that $z_t$ is a function of $(k_t,v_t,x^t)$. The independence of $x_t$ from $(k^{t-1},z^{t-1})$ was also used.  In \eqref{eq:app4} we used the secure encoder assumption (independence of $X^t$ from $Z^t$). Therefore we have, 
\begin{align}
 	P(k^{t-1},x_t,z^t) &= P(x_t,z^t)P(k^{t-1}|x_t,z^t)\nl
	&= P(x_t,z^t)P(k^{t-1}|z^t)
\end{align}
and we proved that \eqref{eq:RT_Markov} is satisfied.

\bibliographystyle{IEEEtran}
\bibliography{PhDBib-1}

\begin{thebibliography}{10}
\providecommand{\url}[1]{#1}
\csname url@samestyle\endcsname
\providecommand{\newblock}{\relax}
\providecommand{\bibinfo}[2]{#2}
\providecommand{\BIBentrySTDinterwordspacing}{\spaceskip=0pt\relax}
\providecommand{\BIBentryALTinterwordstretchfactor}{4}
\providecommand{\BIBentryALTinterwordspacing}{\spaceskip=\fontdimen2\font plus
\BIBentryALTinterwordstretchfactor\fontdimen3\font minus
  \fontdimen4\font\relax}
\providecommand{\BIBforeignlanguage}[2]{{%
\expandafter\ifx\csname l@#1\endcsname\relax
\typeout{** WARNING: IEEEtran.bst: No hyphenation pattern has been}%
\typeout{** loaded for the language `#1'. Using the pattern for}%
\typeout{** the default language instead.}%
\else
\language=\csname l@#1\endcsname
\fi
#2}}
\providecommand{\BIBdecl}{\relax}
\BIBdecl

\bibitem{NeuhoffGilbert1982}
D.~Neuhoff and R.~K. Gilbert, ``Causal source codes,'' \emph{IEEE Transactions
  on Information Theory}, vol.~28, no.~5, pp. 701--713, September 1982.

\bibitem{Shannon1949}
C.~E. Shannon, ``Communication theory of secrecy systems,'' \emph{Bell Systems
  Technical Jouranl}, vol.~28, no.~4, pp. 656--715, 1949.

\bibitem{Wyner1975}
A.~D. Wyner, ``The wire--tap channel,'' \emph{BSTJ}, vol.~54, no.~8, pp.
  1355--1387, 1975.

\bibitem{Yamamoto1997}
H.~Yamamoto, ``Rate-distortion theory for the shannon cipher system,''
  \emph{IEEE Transactions on Information Theory}, vol.~43, no.~3, pp. 827--835,
  May 1997.

\bibitem{Prab-Ramch1997}
V.~Prabhakaran and K.~Ramchandran, ``On secure distributed source coding,'' in
  \emph{Information Theory Workshop, 2007. ITW '07. IEEE}, sept. 2007, pp. 442
  --447.

\bibitem{SlepianWolf73}
D.~Slepian and J.~Wolf, ``Noiseless coding for correlated information
  sources,'' \emph{IEEE Transactions on Information Theory}, vol.~19, pp.
  471--480, 1973.

\bibitem{GunduzEkripPoor2008}
D.~Gunduz, E.~Erkip, and H.~Poor, ``Secure lossless compression with side
  information,'' in \emph{Information Theory Workshop, 2008. ITW '08. IEEE},
  may 2008, pp. 169 --173.

\bibitem{VillardPaint2011}
J.~Villard and P.~Piantanida, ``Secure multiterminal source coding with side
  information at the eavesdropper,'' \emph{CoRR}, vol. abs/1105.1658, 2011.

\bibitem{Merhav2008}
N.~Merhav, ``Shannon's secrecy system with informed receivers and its
  application to systematic coding for wiretapped channels,'' \emph{IEEE
  Transactions on Information Theory}, vol.~54, no.~6, pp. 2723 --2734, June
  2008.

\bibitem{SchielerCuff2013}
C.~Schieler and P.~Cuff, ``Rate-distortion theory for secrecy systems,''
  \emph{CoRR}, vol. abs/1305.3905, 2013.

\bibitem{UduwerelleHoISIT2012}
C.~Uduwerelle, S.-W. Ho, and T.~Chan, ``Design of error-free perfect secrecy
  system by prefix codes and partition codes,'' in \emph{Proc. 2012 IEEE
  International Symposium on Information Theory}, Cambridge, MA, USA, July
  2012, pp. 1593--1597.

\bibitem{Me-ISIT2012}
Y.~Kaspi and N.~Merhav, ``On real-time and causal secure source coding,'' in
  \emph{Proc. 2012 IEEE International Symposium on Information Theory},
  Cambridge, MA, USA, July 2012, pp. 353--357.

\bibitem{TsachyNeri2005}
T.~Weissman and N.~Merhav, ``On causal source codes with side information,''
  \emph{IEEE Transactions on Information Theory}, vol.~51, no.~11, pp.
  4003--4013, November 2005.

\bibitem{Me-RTRD2013}
Y.~Kaspi and N.~Merhav, ``Zero-delay and causal single-user and multi-user
  lossy source coding with decoder side information,'' \emph{{Submitted to}
  IEEE Transactions on Information Theory 2013, CoRR}, vol. abs/1301.0079,
  2013.

\bibitem{AlonOrlitsky1994}
N.~Alon and A.~Orlitsky, ``A lower bound on the expected length of one-to-one
  codes,'' \emph{Information Theory, IEEE Transactions on}, vol.~40, no.~5, pp.
  1670--1672, 1994.

\bibitem{AlonOrlitski96}
------, ``Source coding and graph entropies,'' \emph{IEEE Transactions on
  Information Theory}, vol.~42, no.~5, pp. 1329--1339, September 1996.

\bibitem{DurrettBook}
R.~Durrett and R.~Durrett, \emph{Probability: Theory and Examples}, ser.
  Cambridge Series in Statistical and Probabilistic Mathematics.\hskip 1em plus
  0.5em minus 0.4em\relax Cambridge University Press, 2010.

\bibitem{cover}
T.~M. Cover and J.~A. Thomas, \emph{Elements of Information Theory},
  2nd~ed.\hskip 1em plus 0.5em minus 0.4em\relax Wiley, 2006.

\bibitem{JohnsonIshwarSP2004}
M.~Johnson, P.~Ishwar, V.~Prabhakaran, D.~Schonberg, and K.~Ramchandran, ``On
  compressing encrypted data,'' \emph{Signal Processing, IEEE Transactions on},
  vol.~52, no.~10, pp. 2992--3006, 2004.

\end{thebibliography}
\end{document}